\documentclass{article}
\usepackage{amsfonts}
\usepackage{epsfig}
\usepackage{graphicx}
\usepackage{amssymb}

\title{QUANTIZED KEPLER-COULOMB DYNAMICAL MODELS ON TWO-DIMENSIONAL CONSTANT CURVATURE SPACES}
\author{AGNIESZKA MARTENS\\
Helena Chodkowska  University of Technology and Economics\\
Jutrzenki 135, 02-231  Warszawa, Poland\\
}

\begin{document}

\maketitle
\begin{abstract}
The paper is continuation of \cite{m6} where we have discussed some classical and quantization problems of rigid bodies of infinitesimal size moving in Riemannian spaces. 
Strictly speaking, we have considered oscillatory dynamical models on sphere and pseudosphere.  Here we concentrate on Kepler-Coulomb potential models.
We have used formulated in \cite{m6} the two-dimensional situation on the quantum level. The  Sommerfeld  polynomial  method  is  used  to  perform  the  quantization  of  such  problems. 
The  quantization   of   two-dimensional   problems   may   have   something   to   do   with the   dynamics   of   graphens,   fullerens   and   nanotubes.  This   problem   is also   nearly   related 
 to   the   so-called   restricted   problems   of   rigid   body   dynamic \cite{2}, \cite{JJS}. 
\end{abstract}

\section{Bertrand systems }  

The classical  Bertrand systems are those concerning the motion of material point in the 
spherically-symmetric potential force. It was shown very many years ago that there are two Bertrand 
potentials in Euclidean space for which all bounded orbits are closed: harmonic oscillator and attractive Coulomb problem.
Incidentally, it turned out that it is the peculiarity of dimension three that the attractive Coulomb problem, i.e.,
inverse-square rule for the potential is also the Green function for the Laplace-Beltrami operator.  A very inventive proof of the Bertrand theorem 
may be found, e.g., in the Arnold book  \cite{2}.

There is an interesting feature of our Bertrand problems in  curved manifolds 
visible within the framework of the Hamilton-Jacobi theory. Namely, the energy turns out to be a sum of the geodetic terms 
on the manifolds and of the "usual" terms characteristic for the spherical and pseudospherical geometry.
This opens the possibility of strange conjectures concerning some "experimental" attempts of deciding if the Universe is closed or infinite. 
As usual when dealing with highly symmetric models, there is a parallelism between classical, quasiclassical and quantum theories.
This is to be discussed in the forthcoming paper. Here we concentate on the two-dimensional situation on the quantum level. 

\section{The quantized problems}

As usual when quantizing systems in Riemannian configuration spaces, we shall fix some notation.
Let $Q$ be a differential manifold of dimension $n$
with the metric tensor $G$. The components of $G$ with respect to some local coordinates
$q^{1}, \dots, q^{n}$ will be denoted by $G_{ij}$ and the components of the
contravariant inverse of $G$ will be denoted by $G^{ij}$; obviously, 
$G_{ik}G^{kj}=\delta_{i}{}^{j}$. The determinant of the matrix $[G_{ij}]$ will be
briefly denoted by the symbol $|G|$ (no confusion between two its meanings);
obviously, this determinant is an analytic representation of some scalar 
density of weight two; the square root $\sqrt{|G|}$ is a scalar density of weight one.
The invariant measure induced by $G$ will be denoted by $\widetilde{\mu}$; analytically
its element is given by
\[
d\widetilde{\mu}(q)=\sqrt{|G(q)|}dq^{1} \cdots dq^{n}.
\]
Operators of the covariant differentation induced in the Levi-Civita sense by $G$
will be denoted by $\nabla_{i}$. The corresponding Laplace-Beltrami operator $\Delta$ is 
analytically given by
\[
\Delta=G^{ij}\nabla_{i}\nabla_{j}
\]
or explicitly, when acting on scalar fields,
\[
\Delta{\bf\Psi}=\frac{1}{\sqrt{|G|}}\sum_{i,j}\frac{\partial}{\partial q^{i}}
\left(\sqrt{|G|}G^{ij}\frac{\partial {\bf\Psi}}{\partial q^{j}}\right),
\]
${\bf\Psi}$ denoting a twice differentiable complex function on $Q$.

Wave mechanics is formulated in $L^{2}(Q, \widetilde{\mu})$, the space
of square-integrable functions on $Q$ with the scalar product meant as follows:
\[
\langle{\bf\Psi}|{\bf\Phi}\rangle:=\int \overline{{\bf\Psi}}(q){\bf\Phi}(q)d\widetilde{\mu}(q).
\]

The operator $\Delta$ is symmetric with respect to this product, and $\nabla_{i}$
are skew-symmetric. The metric $G$ underlies the classical kinetic energy, therefore, 
the classical energy/Hamiltonian function 
\[
H=\frac{\mu}{2}G_{ij}(q)\frac{dq^{i}}{dt}\frac{dq^{j}}{dt}+V(q)=\frac{1}{2\mu}
G^{ij}(q)p_{i}p_{j}+V(q)
\]
becomes the operator
\[
\widehat{H}=-\frac{\hbar}{2\mu}\Delta +V.
\]
The Laplace-Beltrami operator depending on the considered manifold has the following form \cite{m6}:

\begin{itemize}
\item[$(i)$] sphere:
\begin{eqnarray}
\Delta & =&\frac{\partial ^{2} }{\partial r ^{2}}+\frac{1}{R}
\cot\frac{r}{R}\frac{\partial }{\partial r }-\frac{2\cos\frac{r}{R}}{R ^{2}\sin
^{2}\frac{r}{R}}\frac{\partial ^{2}}{\partial \varphi \partial \psi} \nonumber \\
&+&\frac{mR^{2}\sin
^{2}\frac{r}{R}+I\cos
^{2}\frac{r}{R}}{IR^{2}\sin
^{2}\frac{r}{R}}\frac{\partial ^{2} }{\partial \psi ^{2}}+\frac{1}{R^{2}\sin
^{2}\frac{r}{R}}\frac{\partial ^{2} }{\partial \varphi ^{2}},\label{EQ4}
\end{eqnarray}
\end{itemize}

\begin{itemize}
\item[$(ii)$] pseudosphere:
\begin{eqnarray}
\Delta &=&\frac{\partial ^{2} }{\partial r ^{2}}+\frac{1}{R}
\coth\frac{r}{R}\frac{\partial }{\partial r }-\frac{2\cosh\frac{r}{R}}{R ^{2}\sinh
^{2}\frac{r}{R}}\frac{\partial ^{2}}{\partial \varphi \partial \psi} \nonumber \\
&+& \frac{ mR^{2}\sinh
^{2}\frac{r}{R}+I\cosh
^{2}\frac{r}{R}}{IR^{2}\sinh
^{2}\frac{r}{R}}\frac{\partial ^{2} }{\partial \psi ^{2}}+\frac{1}{R^{2}\sinh
^{2}\frac{r}{R}}\frac{\partial ^{2} }{\partial \varphi ^{2}}.\label{EQ5a}
\end{eqnarray}
\end{itemize}

The basis of solutions of the stationary Schr\"{o}dinger equation
\[
 \hat{H}\Psi =E\Psi
\]
has the form:
\begin{equation}\label{EQ6}
\Psi(r, \varphi, \psi)=f_{r}(r)f_{\varphi} (\varphi)f_{\psi} (\psi ).
\end{equation}
It is convenient to use the variable $\vartheta=r/R$ for our calculations, then
\begin{equation}\label{EQ6a}
\Psi(\vartheta, \varphi, \psi)=f_{\vartheta}(\vartheta)e^{in\varphi}e^{il\psi},
\end{equation}
where $n, l $ are integers. 

The stationary Schr\"{o}dinger equation with an arbitrary potential
$V(\vartheta)$ leads after the standard separation procedure
to the following one-dimensional radial eigenequations:
\begin{itemize}
\item[$(i)$] sphere:
\[
\frac{d^{2}f_{\vartheta}(\vartheta)}{d\vartheta^{2}}+\cot\vartheta \frac{df_{\vartheta}(\vartheta)}{d\vartheta}-
\]
\[
\left(\frac{\left(\frac{m}{I} R^{2}\sin^{2}\vartheta+\cos^{2}\vartheta\right)n^{2}+l^{2}-2nl\cos\vartheta}{\sin^{2}\vartheta}-
\frac{2mR^{2}}{\hbar^{2}}(E-V(\vartheta))\right)\] 
\[
f_{\vartheta}(\vartheta)=0,
\]

\item[$(ii)$] pseudosphere:
\[
\frac{d^{2}f_{\vartheta}(\vartheta)}{d\vartheta^{2}}+\coth\vartheta \frac{df_{\vartheta}(\vartheta)}{d\vartheta}-
\]
\[
\left(\frac{\left(\pm\frac{ m}{I} R^{2}\sinh^{2}\vartheta+\cosh^{2}\vartheta\right)n^{2}+l^{2}-2nl\cosh\vartheta}{\sinh^{2}\vartheta}-
\frac{2mR^{2}}{\hbar^{2}}(E-V(\vartheta))\right)
\]
\[
f_{\vartheta}(\vartheta)=0.
\]

\end{itemize}

Here we consider the Kepler-Coulomb potential models discussed in \cite{m6}. Then, the resulting
Schr\"{o}dinger equations may be rigorously solvable in terms of some standard
special functions. The Sommerfeld polynomial
method is used to perform the quantization of such problems
\cite{m5}-\cite{m7}. The solutions are expressed by 
the usual or confluent Riemann $P$-functions, which are deeply related to the 
hypergeometric functions. Supposing that the usual
convergence demands are imposed, then the hypergeometric functions become polynomials 
and the solutions are expressed by elementary functions. The energy 
levels  are expressed by the eigenvalues of the corresponding
operators. We restrict ourselves to the solutions expressible in terms of Riemann 
$P$-functions  because this class of functions is well investigated moreover
many special functions used in physics may be expressed by them.

\section{Examples}  

The above equations may be
solved only when the explicit form of potential is specified. 
Considered is a special case, when the translational part of the potential energy
$V(\vartheta)$ ($V(r)$) has the Bertrand structure, 
i.e. with the "frozen" rotations all orbits would be closed \cite{m6}
\begin{itemize}
\item[$(i)$] sphere:
\end{itemize}
Here we consider the following model of the  Kepler-Coulomb potentials:
\begin{equation}\label{c88}
V(r)=-\frac{\alpha}{R}{\rm \cot}\frac{r}{R}.
\end{equation}
Applying the Sommerfeld polynomial method we obtain the energy
levels $E$ as follows:\\
\begin{equation}\label{EQ8}
E=-\frac{\hbar^{2}}{2mR^{2}}\left( (k+|n+l||n-l|) -\left( \left(\frac{m}{I}R^{2}+1 \right)n^{2}-4nl-2\alpha \right) \right),
\end{equation}
where $k=0,1,... \ $. After some calculations we
obtain the function $f_{r}(r)$ in the form:
\begin{equation}
f_{r}(r)=\left(\sin \frac{r}{R}\right)^{\kappa}\left(\cos \frac{r}{R}\right)^{\nu}  F\left(-k,k+\kappa+\nu;1+\kappa;\sin \frac{r}{R}\right),
\end{equation}
where 
\[
\kappa=|n+l|, \quad \nu=|n-l|.
\]
\begin{itemize}
\item[$(ii)$] pseudosphere:
\end{itemize}
We take the "attractive Kepler-Coulomb" - type potentials: 
\begin{equation}\label{b133}
 V(r)=-\frac{\alpha}{R} {\rm \coth}\frac{r}{R}, \qquad \alpha>0.
\end{equation}
\bigskip
We find the energy levels $E$ in the form:
\begin{equation}\label{EQ9}
E=-\frac{\hbar^{2}}{2mR^{2}}\left( (k+|n+l||n-l|) +\left( \left(\pm \frac{m}{I}R^{2}+1 \right)n^{2}-4nl-2\alpha \right) \right).
\end{equation}
The function $f_{r}(r)$ is as follows:
\begin{equation}
f_{r}(r)=\left(\sinh \frac{r}{R}\right)^{\kappa}\left(\cosh \frac{r}{R}\right)^{\nu}  F\left(-k,k+\kappa+\nu;1+\kappa;\sinh \frac{r}{R}\right).
\end{equation}

\section{Conclusions}

The considered Kepler-Coulomb dynamical models are completely non-degenerate. 
This fact is reflected by the existence of three
quantum numbers labelling the energy levels, which cannot be               
combined into a single quantum number, i.e., there is no total
quantum degeneracy (hyperintegrability) with respect to them.
We can notice that, the  interaction  between translational  and  rotational  degrees  of  freedom  completely  removes  degeneracy.
As yet it is not clear for us if some weaker degeneracy does occur for some relationships between  constants $m$, $I$, $R$, $\alpha$. 
Later on, we shall consider test affinelly-rigid body \cite{m7} and corresponding quantized Kepler-Coulomb potential models.


\begin{thebibliography}{99}

\bibitem{2} V. I. Arnold:  {\em Mathematical Methods of Classical Mechanics\/}, Springer Graduate Texts in Mechanics {\bf 60}, Springer Verlag, New York 1978.

\bibitem{m4}
A. Martens:  {\em J.\ Nonlinear Math.\ Phys.} {\bf 11},   Suppllement, 145  (2004). 

\bibitem{m5}
A. Martens:  {\em J.\ Nonlinear Math.\ Phys. }{\bf 11},   Suppllement, 151  (2004).

\bibitem{AM} A. Martens:  {\em Rep. Math. Phys.} {\bf 62}, 145  (2008).

\bibitem{m2} A. Martens, J. J. S\l awianowski:  {\em Acta\ Phys.\ Pol. B\/} {\bf 41}, 1847  (2010).

\bibitem{m6}
A. Martens:  {\em Rep. Math. Phys.} {\bf 71}, 381  (2013).

\bibitem{m7}
A. Martens:  {\em Acta\ Phys.\ Pol. B\/}  {\bf 46},  843  (2015).

\bibitem{JJS} J.J. S\l awianowski: {\em Geometry of Phase Spaces},  John Wiley \& Sons, Chichester, New York, Brisbane, Toronto, Singapore, PWN-Polish Scientific Publishers, Warszawa 1991. 



\end{thebibliography}
\end{document}